\documentclass[paper]{JHEP3}
\usepackage{amssymb,enumerate}
\usepackage{amsmath}
\usepackage{bbm}
\usepackage{url}
\usepackage{cite}
\usepackage{graphics}
\usepackage{xspace}
\usepackage{epsfig}
\usepackage{subfigure}

\newcommand\POWHEG{{\tt POWHEG}}
\newcommand\POWHEGBOX{{\tt POWHEG~BOX}}
\newcommand\PYTHIA{{\tt PYTHIA}}

\def\({\left(} 
\def\){\right)} 

\def\beq{\begin{equation}}
\def\beqn{\begin{eqnarray}}
\def\eeq{\end{equation}}
\def\eeqn{\end{eqnarray}}

\def\mr{\mathrm}

\def\vbfz{VBF~$Zjj$\;}
\def\lpm{\ell^+\ell^-}
\def\lljj{\ell^+\ell^-jj}
\def\eejj{e^+e^-jj}

\def\muf{\mu_F}
\def\mur{\mu_R}

\title{Next-to-leading order QCD corrections to electroweak $Zjj$ production in the \POWHEGBOX} 
\vfill
\author{
Barbara J\"ager \\
Institut f\"ur Physik (THEP), Johannes-Gutenberg-Universit\"at, 55099 Mainz, Germany\\
E-mail: \email{jaegerba@uni-mainz.de}
}
\author{
Steven Schneider \\
Institut f\"ur Physik (THEP), Johannes-Gutenberg-Universit\"at, 55099 Mainz, Germany\\
E-mail: \email{schneste@students.uni-mainz.de}
}

\author{Giulia Zanderighi \\
Rudolf Peierls Centre for Theoretical Physics, 1 Keble Road, University of Oxford, UK\\
E-mail: \email{g.zanderighi1@physics.ox.ac.uk}

}


\abstract{We present an implementation of electroweak $Z$-boson
  production in association with two jets at hadron colliders in the
  \POWHEG{} framework, a method that allows the interfacing of NLO-QCD
  calculations with parton-shower Monte Carlo programs.
We focus on the leptonic decays of the weak gauge boson, and take
photonic and non-resonant contributions to the matrix elements fully
into account.
We provide results for observables of particular importance for the
suppression of QCD backgrounds to vector-boson fusion processes by
means of central-jet-veto techniques. While parton-shower effects are
small for most observables associated with the two hardest jets, they
can be more pronounced for distributions that are employed in
central-jet-veto studies.

}
\preprint{MZ-TH/12-25\\
  OUTP-12-13P}

\begin{document}

\section{Introduction}
One of the most central themes of the CERN Large Hadron Collider (LHC)
is the exploration of the mechanism of electroweak symmetry breaking,
which in the context of the Standard Model proceeds via a CP-even
spin-zero particle, the Higgs boson. The discovery of a new particle
that is compatible with the postulated Higgs boson \cite{higgs-2012}
is thus a breakthrough in our understanding of the electroweak
interactions. However, to truly establish the existence of a {\em
  Standard-Model like} Higgs boson, besides its mass a variety of
additional properties, such as its decay widths, couplings to gauge
bosons and fermions, its spin and CP quantum numbers, have to be
determined and confronted with theory predictions. This can only be
achieved if background processes that are omni-present at hadron
colliders are well under control.

A reaction that has already been exploited for the postulated Higgs
discovery by the {\tt ATLAS}~\cite{atlas-higgs-2012} and {\tt
  CMS}~\cite{cms-higgs-2012} collaborations, and that will play an
even more important role in the upcoming determination of the new
particle's properties is Higgs production via weak vector boson fusion
(VBF), i.~e.\ the purely electroweak $qq\to qqH$ process that proceeds
via weak-boson exchange in the $t$-channel. Because of the color
singlet nature of this exchange, VBF processes feature two jets in the
forward regions of the detector and little jet activity at central
rapidities, while decay products of the Higgs boson tend to be located
in between the two tagging jets.  These characteristic properties can
be exploited to distinguish VBF processes from a priori overwhelming
QCD backgrounds.

To explore these features on the basis of a Standard-Model reaction
with very similar properties as VBF Higgs production, in
Ref.~\cite{Rainwater:1996ud} it has been suggested to consider the
related process of electroweak $Zjj$ production as a testing ground
for the study of $t$-channel color-singlet exchange events.  Based on
a leading order (LO) parton-level calculation for the electroweak (EW)
$qq\to qqZ$ process, supplemented by explicit matrix elements with an
extra parton in the final state for the simulation of additional jet
activity, it was shown that color-singlet exchange in the $t$-channel
gives rise to soft minijet activity that differs considerably from
that of QCD-initiated background processes. In particular, imposing a
{\em central jet veto} (CJV), i.~e.\ discarding events with hard jets
at central rapidities, substantially improves the signal significance
of VBF processes.

However, in order to reliably predict CJV efficiencies, a precise
knowledge of signal and background processes is essential.
Within inclusive selection cuts, the dominant source of $Zjj$ events
at hadron colliders is QCD-induced production. Next-to-leading order
(NLO) QCD corrections to this process are available in the form of a
flexible parton-level Monte-Carlo program in the public {\tt
  MCFM}~package~\cite{Campbell:2002tg,Campbell:2003hd}. The virtual
corrections of this implementation have been adapted from
Ref.~\cite{Bern:1997sc,Nagy:1998bb}, while the real-emission
contributions have first been provided in
\cite{Berends:1988yn,Hagiwara:1988pp}.
A merging of the NLO-QCD corrections to QCD-induced $Zjj$ production
with parton-shower programs has been provided in
Ref.~\cite{Re:2012zi}.
In an inclusive setup, the electroweak $Zjj$ production cross section
is much smaller than the QCD induced once. However, once VBF-specific
selection cuts are applied, the signature of the electroweak
production mode is quite distinct and can thus be considered
separately.
The NLO-QCD corrections to electroweak $Zjj$ production, including
``signal-type'' VBF diagrams, but also $Z/\gamma^\star$ bremsstrahlung
and non-resonant contributions, were computed in
Ref.~\cite{Oleari:2003tc}.
They are available in the public {\tt VBFNLO} program package
\cite{Arnold:2008rz}. However, an interface of this code to a
parton-shower program (PS) such as {\tt
  HERWIG}~\cite{Marchesini:1991ch,Corcella:2000bw} or {\tt
  PYTHIA}~\cite{Sjostrand:2006za} is not available to date.

In this work, we aim at providing this as yet missing tool by
implementing an NLO-QCD calculation for electroweak $Zjj$ production
in the so-called \POWHEGBOX{}\cite{Alioli:2010xd}, a framework that
allows to match dedicated NLO-QCD calculations with public
parton-shower programs as described in some detail, e.~g.~in
Refs.~\cite{Nason:2004rx,Frixione:2007vw}.  After an outline of the
technical aspects of this endeavor in Sec.~\ref{sec:tech}, we provide
phenomenological results in Sec.~\ref{sec:pheno}. The theoretical
uncertainties associated with our calculation are discussed and the
impact of parton-shower effects on observables that are utilized in
CJV studies is illustrated. Our conclusions are given in
Sec.~\ref{sec:conc}.

\section{Technical details}
\label{sec:tech}
For developing an interface between a parton-level calculation for
VBF-induced $Zjj$ production at NLO-QCD accuracy and parton-shower
programs, we have made use of the publicly available
\POWHEGBOX{}~\cite{Alioli:2010xd}. This package contains all
process-independent building blocks needed for the matching of a
dedicated NLO-QCD calculation in the context of the \POWHEG{}
framework~\cite{Nason:2004rx,Frixione:2007vw} with multi-purpose
parton-shower Monte-Carlo programs. Process-specific ingredients have
to be provided by the user of the \POWHEGBOX{}. These include in
particular:
\begin{itemize}
\item
a list of all flavor structures contributing to the Born process, 
\item
the Born phase space, 
\item the Born amplitudes squared for all partonic subprocesses and
  the color correlated and the spin correlated Born amplitudes,
\item the Born color structure in the limit of a large number of colors, 
\item the finite part of the virtual corrections,
\item a list of all flavor structures contributing to the
  real-emission process,
\item the real-emission matrix elements squared for all partonic
  subprocesses.
\end{itemize}
Once these building blocks have been implemented, the \POWHEGBOX{}
itself takes care of infrared singularities by means of an FKS-type
subtraction procedure~\cite{Frixione:1995ms}.

The user can choose to run the program in a parton-level mode at LO or
NLO-QCD, which is particularly useful for validation purposes. In
addition, the \POWHEGBOX{} provides the interface to any $p_T$-ordered
parton-shower program such as {\tt PYTHIA}. Transverse momentum
ordering is essential, since the \POWHEG{} method relies on generating
the hardest emission in an event first, while subsequent emissions
have to be provided by the parton-shower program.
The \POWHEGBOX{} can also be matched with angular-ordered parton
shower programs, such as {\tt HERWIG}, if a so-called vetoed-truncated
shower is provided.  The public version of {\tt HERWIG} does not offer
this option, however, and results obtained by combining an NLO-QCD
calculation via the \POWHEGBOX{} with {\tt HERWIG} can thus only be
accurate up to small effects of this missing vetoed-truncated
shower. For our phenomenological analysis, we will therefore restrict
ourselves to \POWHEG{} matched with {\tt PYTHIA}.

In the course of the last few years, a variety of Standard-Model
processes involving jets in the final state has successfully been
implemented in the \POWHEGBOX{}, including dijet
production~\cite{Alioli:2010xa} and QCD and EW
$Hjj$~\cite{Campbell:2012am,Nason:2009ai} and
$W^+W^+jj$~\cite{Melia:2011gk,Jager:2011ms} production. Here, we
follow closely the procedure of Ref.~\cite{Jager:2011ms} for the
$W^+W^+jj$ mode with appropriate modifications related to the reduced
multiplicity and more involved singularity structure of the
electroweak $Zjj$ production process.

\begin{figure}[tp]
\centering
  \subfigure[]{\label{fig:zjj_a}\includegraphics[width=0.4\textwidth]{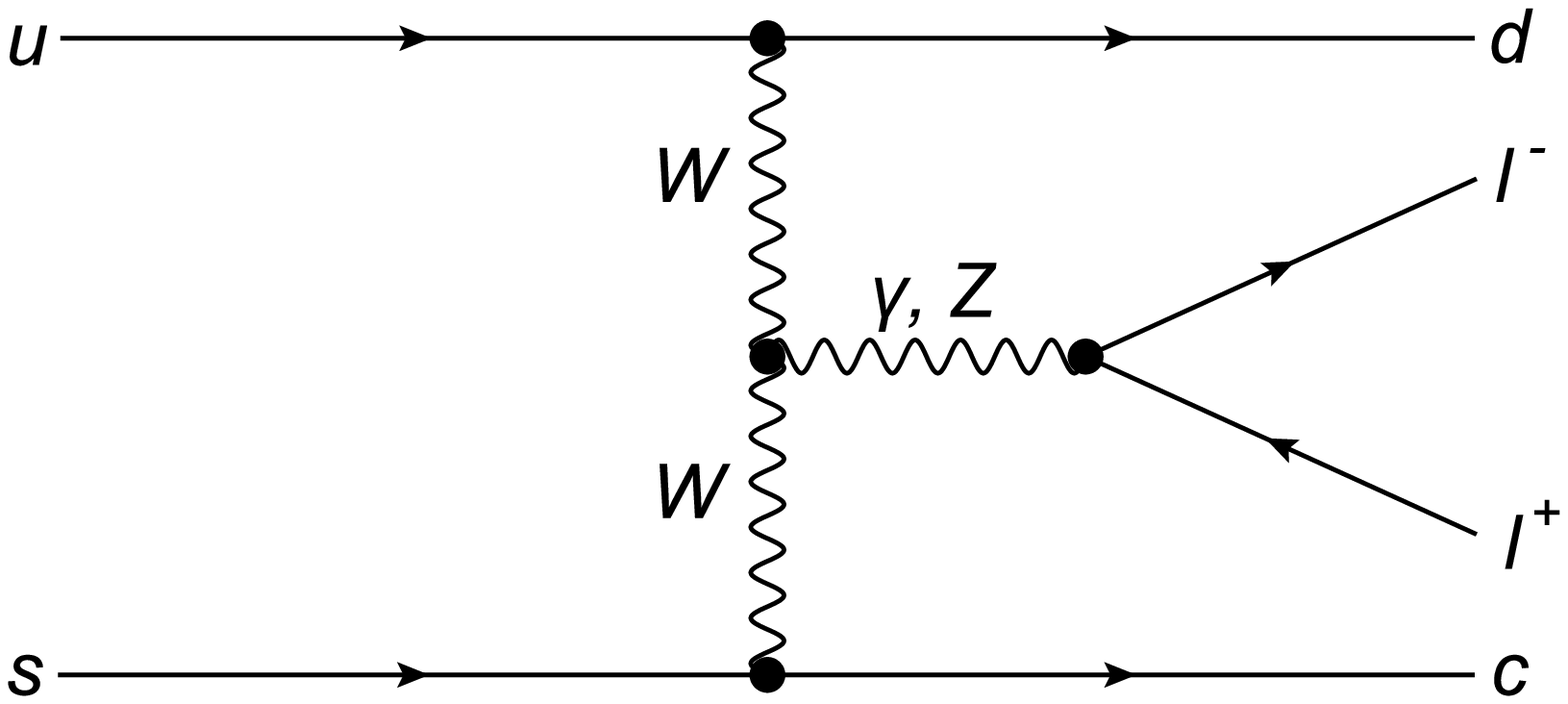}}\hfill
  \subfigure[]{\label{fig:zjj_b}\includegraphics[width=0.4\textwidth]{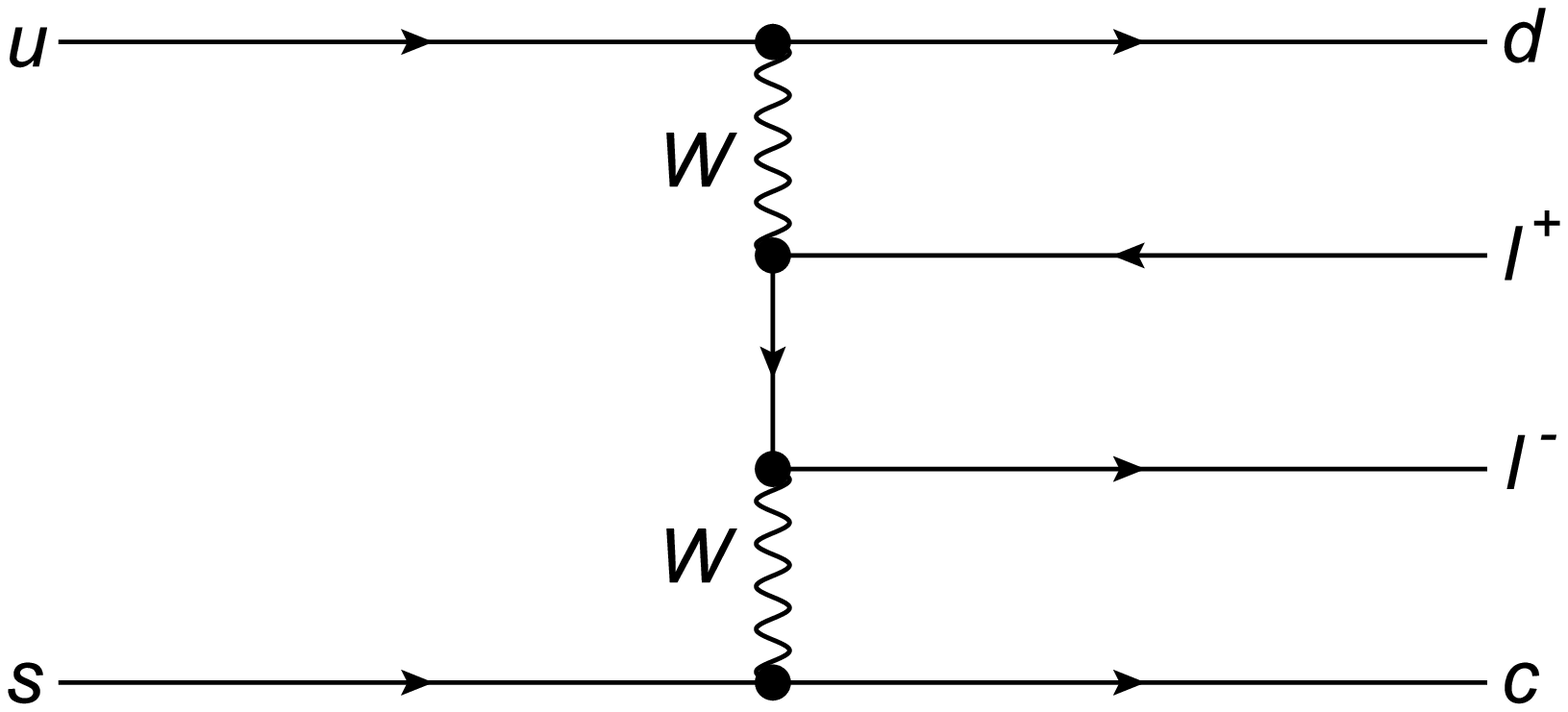}}\hfill
  \subfigure[]{\label{fig:zjj_c}\includegraphics[width=0.4\textwidth]{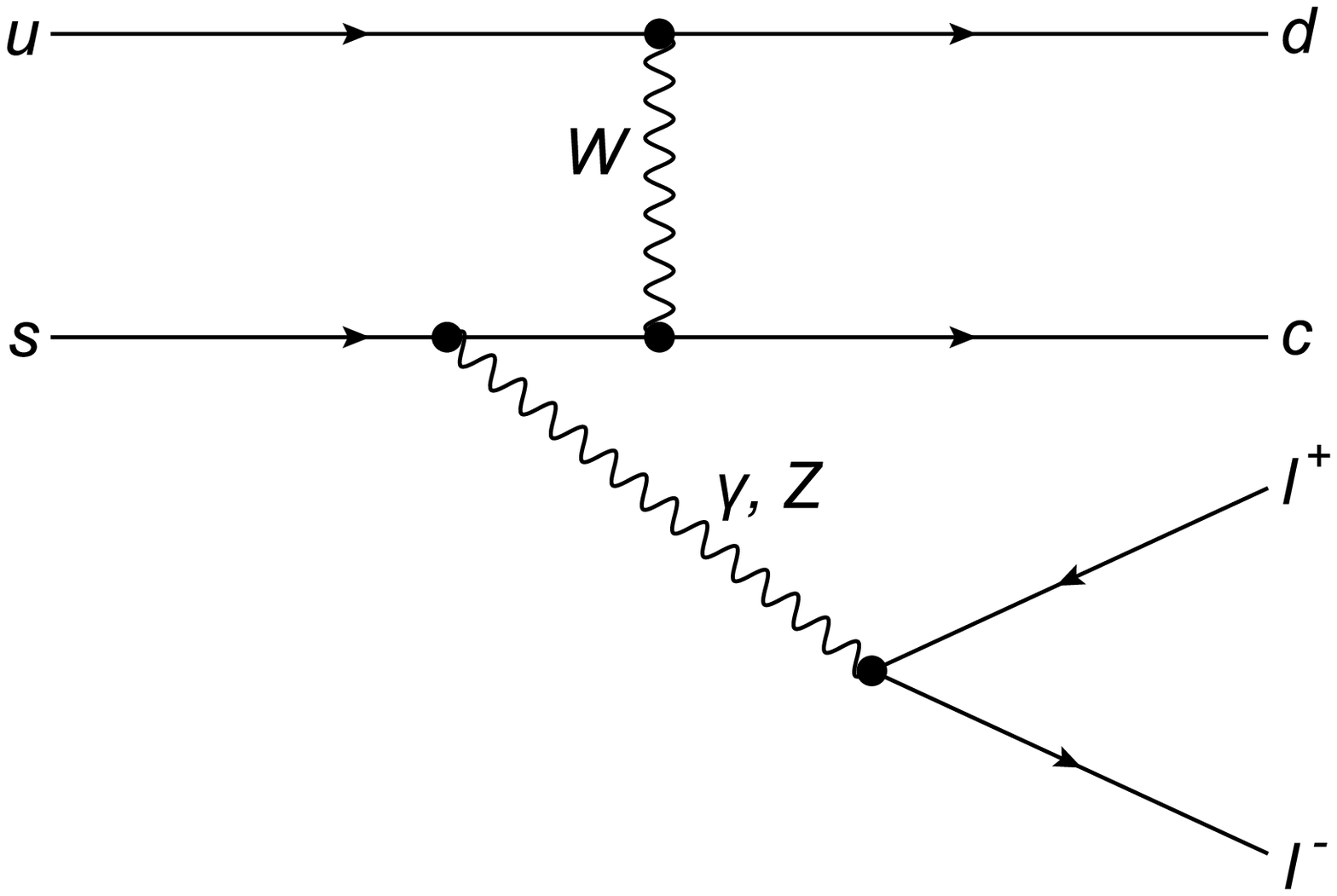}}\hfill
  \subfigure[]{\label{fig:zjj_d}\includegraphics[width=0.4\textwidth]{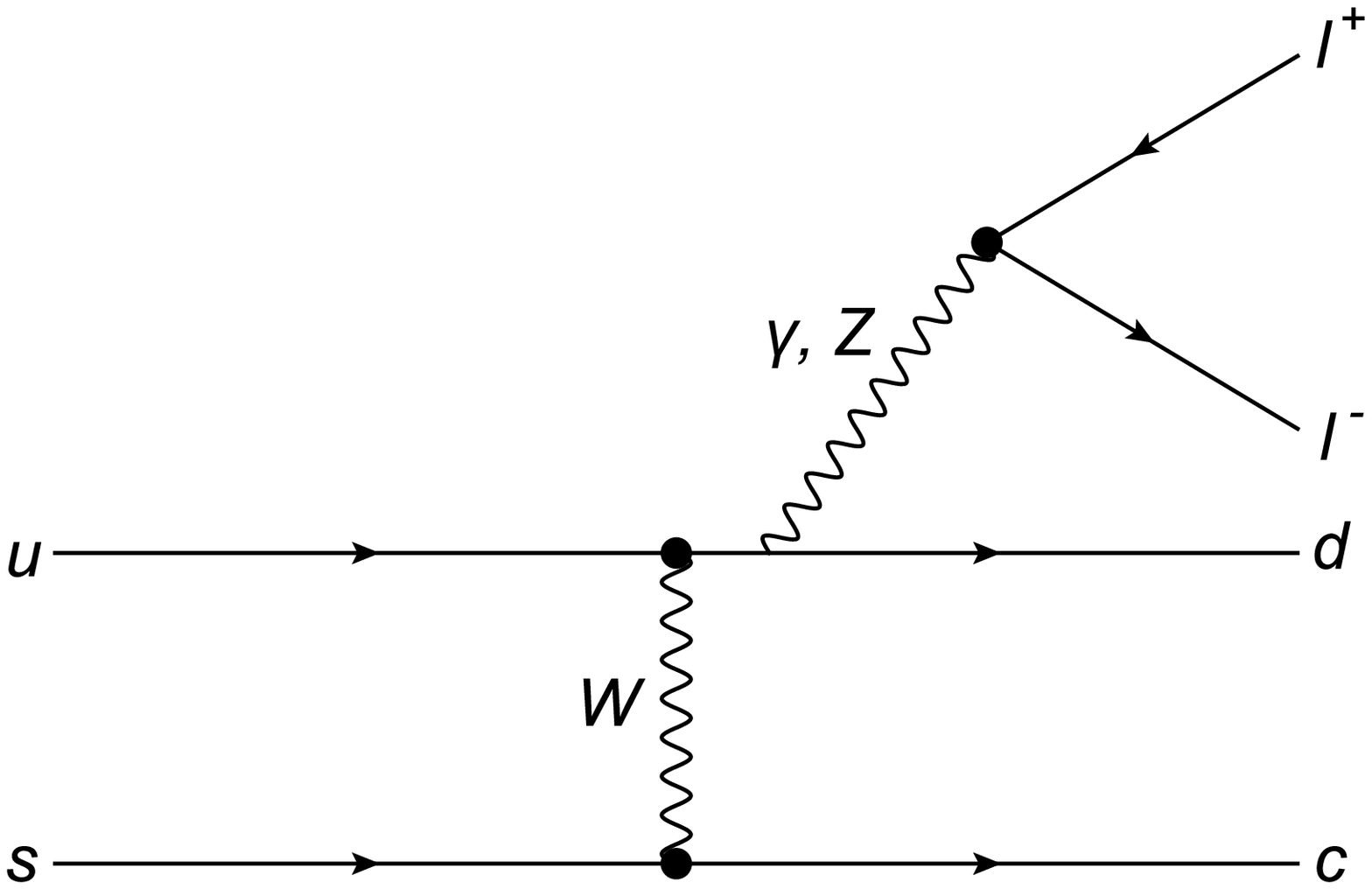}}\hfill
  \caption{ Sample diagrams for the partonic subprocess
    $us\rightarrow\ell^+\ell^-dc$ at leading order.  }
\label{fig:zjj}
\end{figure}

Electroweak $Zjj$ production can proceed via processes of the type
$qq\to qqZ$ with color-singlet $\gamma/Z$ or $W$ exchange, referred to
as neutral current (NC) and charged current (CC) reactions,
respectively. Numerically, the CC contributions dominate over the NC
contributions, mainly because of the larger coupling of the quarks to
$W$~bosons than to $Z$~bosons and photons.
In order to retain all possible angular correlations between leptonic
decay products of the $Z$~boson, we consider EW processes of the type
$q q'\to \ell^+\ell^-\,q\, q'$ (and all related sub-processes with
quarks being replaced by anti-quarks), where $\ell^\pm$ denotes an
$e^\pm$ or a $\mu^\pm$. Since the same final state can be produced via
a virtual photon rather than a $Z$ boson, diagrams with a
$\gamma^\star$ have to be taken into account as well.  Non-resonant
diagrams, where the $\lpm$ pair does not stem from a
$Z/\gamma^\star$-decay but is produced in the $t$-channel have been
included. Some representative Feynman diagrams for the
$us\rightarrow\ell^+\ell^-dc$ subprocess are depicted in
Fig.~\ref{fig:zjj}.
In all diagrams, we are using a fixed width in the $Z$- and $W$-boson
propagators. The uncertainty related to the treatment of massive
gauge-boson propagators is at the order of 0.5\%~\cite{Oleari:2003tc},
and can thus be considered as a minor contribution to higher-order
electroweak corrections.

Note that we do neglect quark-antiquark annihilation diagrams that
contain vector-boson pair production with subsequent decay of one of
the weak bosons into a pair of jets, and interference effects between
$t$-channel and $u$-channel diagrams in subprocesses with identical
quarks. As explained in some detail in Ref.~\cite{Oleari:2003tc}, in
the phase-space regions where VBF processes are searched for, these
contributions are entirely negligible, amounting to no more than 0.3\%
of the full LO results, once VBF-specific selection cuts are applied.
In the absence of selection cuts, at NLO initial-state singularities
can arise from collinear $q\to qg$ and $g\to q\bar q$ splittings. They
are taken care of by being factorized into the respective quark and
gluon distribution functions of the scattering hadrons. Similar
divergences can occur in diagrams where a photon of low virtuality is
exchanged in the $t$-channel, thus giving rise to a collinear $q\to
q\gamma$ configuration. Such contributions are considered as part of
the QCD corrections to $p\gamma\to Zjj$ that we are not providing in
this work. Following the strategy of Ref.~\cite{Oleari:2003tc}, to
avoid singular contributions of this type we impose a cut on the
virtuality of the photon,
$Q^2_{\gamma,\rm{min}}=4$~GeV$^2$. Contributions of lower virtuality
are suppressed by a strong damping factor. We have checked that by
varying the value of $Q^2_{\gamma,\rm{min}}$ in the range from 0.1 to
9~GeV$^2$, the cross section within typical VBF cuts does not change
within the statistical error.
For simplicity we refer to the EW $pp\to \lpm jj$ production process
within the above approximations as ``VBF $Zjj$ production''.

In contrast to Higgs and same-sign gauge boson pair production via
VBF, the integrated cross section for single gauge boson production in
association with two jets is divergent at leading order, unless
dedicated selection cuts are applied. Special care is thus required to
avoid singularities in the generation of the underlying Born
configuration of our simulation.
In order to prevent the population of regions in phase space that are
discarded anyway as soon as realistic selection cuts are applied on
the generated event sample, a so-called Born-suppression factor has
been applied in previous works on $Z$ production processes in
association with one or two jets in the framework of
\POWHEG{}~\cite{Alioli:2010qp,Re:2012zi}.
In \vbfz production, the singular configurations at Born level can be
identified by the transverse momenta of the two final-state partons
and the invariant mass of the leptons $m_{\ell\ell}$. Singular
$\gamma^\star\to\ell^+\ell^-$ configurations are most easily taken
care of by a generation cut on the dilepton invariant mass, e.\ g.\
\beq
\label{eq:gencut-mll}
m_{\ell\ell}^\mr{gen} = 30~\mr{GeV}\,, \eeq
supplemented by a tighter analysis cut on $m_{\ell\ell}$ in our
numerical studies.
The kinematics of the dilepton system is not affected by hadronic
parton-shower effects, making the use of an explicit generation cut on
$m_{\ell\ell}$ unproblematic as long as we disallow QED radiation by
the parton shower.

In addition, we use a Born-suppression factor $F(\Phi_n)$ that
vanishes whenever a singular region of the Born phase space $\Phi_n$
is approached. The \POWHEGBOX{} then generates the underlying Born
kinematics according to a modified $\bar B$ function,
\beq
\bar B_\mr{supp} = \bar B(\Phi_n) F(\Phi_n)\,.
\eeq
Similarly to the prescription of Ref.~\cite{Re:2012zi}, we set
\beq
\label{eq:bsupp}
F(\Phi_n) = 
\left(\frac{p_{T,1}^2}{p_{T,1}^2+\Lambda_{p_T}^2}\right)^{n}
\left(\frac{p_{T,2}^2}{p_{T,2}^2+\Lambda_{p_T}^2}\right)^{n}\,,
\eeq
where the $p_{T,i}$ are the transverse momenta of the two outgoing
partons of the underlying Born configuration, and the $\Lambda_{p_T}$
and $n$ are technical parameters to be set by the user. In regions
that are singular because of the outgoing partons' configuration, the
function $F(\Phi_n)$ approaches zero fast enough to yield a finite
value for $\bar B(\Phi_n) F(\Phi_n)$. The function $\bar B_\mr{supp}$
can therefore be used to generate underlying Born configurations. The
generated events then have to be weighted with an extra factor
$1/F(\Phi_n)$ to compensate for the artificial suppression.
As default in our analysis we are using the Born-suppression factor of
Eq.~({\ref{eq:bsupp}) with $\Lambda_{p_T}=10$~GeV,
  $n=2$.~\footnote{Instead of using a Born-suppression factor, one
    could use suitable generation cuts on the transverse momenta of
 the final-state partons in the Born configuration.}

At NLO-QCD, the interference of the Born amplitudes with one-loop
diagrams and real-emission amplitudes squared have to be
considered. Within our approximations, the virtual corrections
comprise only up to box-type corrections to either the upper or
the lower quark line. Diagrams where a gluon is exchanged between the
two quark lines vanish when interfered with the Born amplitude because
of color conservation. Following the procedure of
Ref.~\cite{Jager:2010aj}, the finite parts of the virtual corrections
have been calculated numerically by a Passarino-Veltman type tensor
reduction.
To the real-emission contributions, diagrams with an extra gluon in
the final state [such as the subprocess $q\,q'\to \ell^+\ell^-\,q\,
q'\, g$] and crossing-related reactions with a gluon in the initial
state [e.~g., $g\,q'\to \ell^+\ell^-\,q\, q'\,\bar q$] contribute. For
the calculation of the respective matrix elements we employ the
helicity-amplitude formalism of Ref.~\cite{Hagiwara:1985yu}.

While infrared singularities in the NLO-QCD contributions have been
treated by means of a Catani-Seymour dipole subtraction
procedure~\cite{Catani:1996vz} in the parton-level Monte-Carlo program
of Ref.~\cite{Oleari:2003tc}, the user of the \POWHEGBOX{} does not
need to provide subtraction terms explicitly. The \POWHEGBOX{} rather
generates itself counterterms that take care of potential
singularities in soft and collinear configurations in the context of
the FKS subtraction procedure~\cite{Frixione:1995ms}, using
process-specific information contained in the partonic matrix
elements. In addition, the program checks automatically that the
real-emission contributions approach their soft and collinear limits
correctly. While this latter test provides a useful handle to verify
the flavor structure and implementation of the real-emission
amplitudes by the user, comparing integrated cross sections at NLO-QCD
accuracy obtained with the \POWHEGBOX{} in the FKS framework with a
stand-alone parton-level code based on the Catani-Seymour dipole
subtraction formalism yields a strong check on the entire set-up of
the code.

The virtual corrections of our implementation have been checked
against the corresponding contributions of Ref.~\cite{Oleari:2003tc}
in the public code~\cite{Arnold:2008rz} at amplitude level. In order
to validate our tree-level and real-emission amplitudes we have
compared them for selected phase-space points to respective amplitudes
generated automatically by the {\tt MadGraph} package
\cite{Stelzer:1994ta,Alwall:2007st}. We found agreement at the level
of 10 significant digits. In addition, we have run the \POWHEGBOX{} in
a parton-level standalone mode at LO and NLO and compared integrated
cross sections as well as a variety of kinematic distributions to an
appropriately adapted version of {\tt VBFNLO}, finding full agreement
within the numerical accuracy of the two programs. This provides a
strong check on the consistent implementation of all LO and NLO matrix
elements as well as on the phase-space integration.

\section{Phenomenological results}
\label{sec:pheno}
The implementation of VBF $Zjj$ production in the \POWHEGBOX{} is
publicly available via the web site of the \POWHEGBOX{} project, {\tt
  http://powhegbox.mib.infn.it}, where also instructions for
downloading the code are provided. With the downloaded code version,
the user can perform studies with her/his own preferred
settings. Recommended values for technical parameters and run-time
estimates are provided in the documentation of the VBF~$Zjj$
code. Here, we wish to present representative results for $pp \to
e^+e^-jj$ obtained with our \POWHEGBOX{} implementation for
VBF-specific settings.

We consider $pp$ collisions at a center-of-mass energy of
$\sqrt{s}=8$~TeV.  As electroweak input parameters we use the mass of
the $Z$~boson, $m_Z=91.188$~GeV, the mass of the $W$~boson,
$m_W=80.419$~GeV, and the Fermi constant, $G_F=1.16639\times
10^{-5}$~GeV$^{-1}$. The other EW parameters are computed thereof via
tree-level electroweak relations. The widths of the weak gauge bosons
are set to $\Gamma_Z=2.51$~GeV and $\Gamma_W=2.099$~GeV,
respectively. We assume a diagonal form of the
Cabibbo-Kobayashi-Maskawa matrix. For the parton distribution
functions of the protons, we use the NLO set of the MSTW2008
parametrization~\cite{Martin:2009iq} as implemented in the {\tt
  LHAPDF} library~\cite{Whalley:2005nh}, corresponding to
$\alpha_s(m_Z)=0.12018$. Contributions with $b$-quarks in the initial
state are not taken into account. Jets are defined according to the
anti-$k_T$ algorithm~\cite{Cacciari:2008gp} as available in the {\tt
  FASTJET} package~\cite{Cacciari:2005hq,Cacciari:2011ma}, with
$R=0.4$.
For our NLO+PS analysis, we will use {\tt PYTHIA 6.4.25}, including
hadronization corrections and underlying event with the {\tt Perugia
  0} tune. QED radiation effects in the shower are switched off.
For the representative results we present below, we set the
factorization and renormalization scales to
\beq
\muf = \mur =  M_Z\,.
\eeq
Of course, the user is free to make a different choice for the scales
when using the code, in particular one could also choose to use
dynamical, local scales as suggested in the approach
of~\cite{Hamilton:2012np}.
For all our analyses we require the presence of at least two jets with 
\beq
\label{eq:pt-vbf-cut}
p_{T,j}=20~\mr{GeV}\,,
\quad 
|y_j| < 4.5\,. 
\eeq
The two hardest jets inside this rapidity region are referred to as
``tagging jets''.  Furthermore, the invariant mass of the charged
lepton pair has to be in a narrow window around the mass of the
$Z$~boson,
\beq
\label{eq:mll-cut}
m_Z-10~\mr{GeV}< m_{\ell\ell}< m_Z+10~\mr{GeV}\,,
\eeq
to avoid contributions from collinear $\gamma^\star\to\ell^+\ell^-$
splittings.
In order to enhance the relative importance of VBF contributions to
$pp\to \lljj$ with respect to potential QCD backgrounds, in addition
to Eqs.~(\ref{eq:pt-vbf-cut}) and (\ref{eq:mll-cut}) we require the
two tagging jets to be well-separated in rapidity, lie in opposite
hemispheres of the detector, and exhibit a large invariant mass,
\beq
|\Delta y_{j_1j_2}| = |y_{j1}-y_{j2}| > 4\,,
\quad
y_{j1}\times y_{j2}< 0\,,
\quad
m_{j_1j_2} > 600~\mr{GeV}\,.
\eeq
We furthermore require two charged leptons with
\beq
p_{T,\ell} > 20~\mr{GeV}\,,\quad
|y_\ell| < 2.5\,, 
\eeq
well-separated from each other and from the tagging jets,
\beq
\Delta R_{\ell\ell}> 0.1\,, \quad
\Delta R_{j\ell}> 0.4\,, 
\eeq
and in the rapidity range between the two tagging jets, 
\beq
\label{eq:rapgap-cut}
\min\{y_{j_1},y_{j_2}\}<y_\ell<\max\{y_{j_1},y_{j_2}\}\,.
\eeq

Figure~\ref{fig:jet2} shows our results for the transverse momentum
and rapidity distributions of the second hardest tagging jet within the VBF
cuts of Eqs.~(\ref{eq:pt-vbf-cut})--(\ref{eq:rapgap-cut}) at NLO-QCD
accuracy and for {\tt POWHEG+PYTHIA}.
%
%
\begin{figure}[tp]
\includegraphics[width=0.5\textwidth]{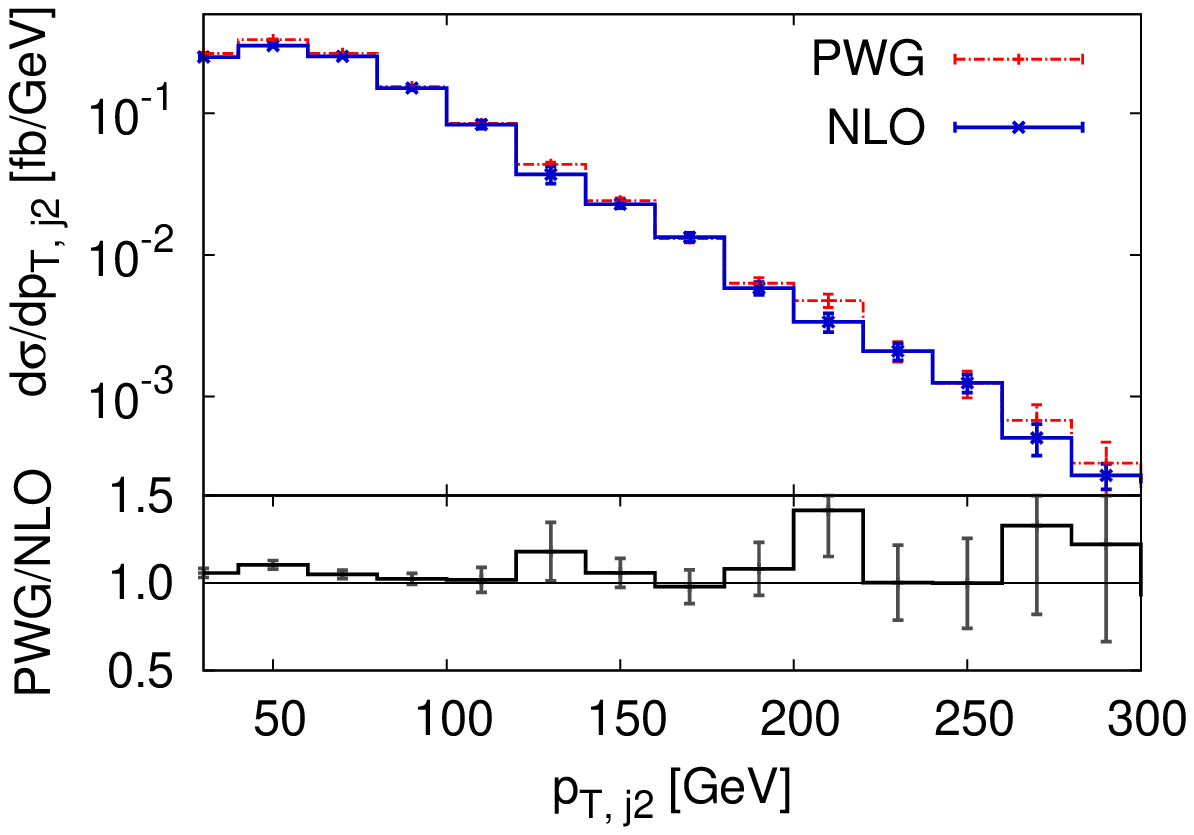}\hfill
\includegraphics[width=0.5\textwidth]{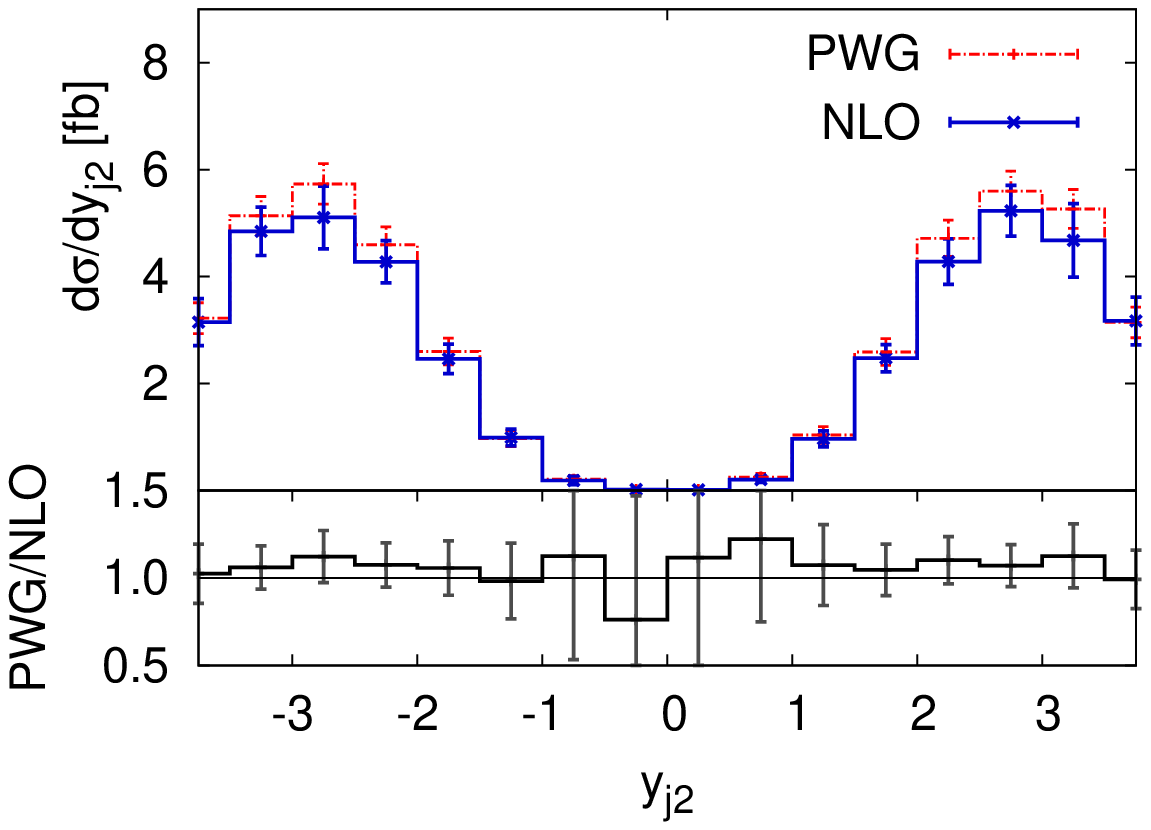}
\caption{
\label{fig:jet2}
Transverse momentum (left panel) and rapidity distribution (right
panel) of the second hardest tagging jet with VBF cuts of
Eqs.~(\ref{eq:pt-vbf-cut})--(\ref{eq:rapgap-cut}) at NLO-QCD ('NLO',
solid blue lines) and with {\tt POWHEG+PYTHIA} ('PWG', dashed red
lines)) for $\eejj$~production at the LHC with $\sqrt{s}=8$~TeV. The
lower panels show the respective ratios of the {\tt POWHEG+PYTHIA} to
the NLO results.  }
\end{figure}
The shapes of the NLO curves for these observables do not change
considerably when the fixed-order calculation is combined with
\PYTHIA{}.  The difference between the normalization of the respective
NLO and NLO+PS results can be traced back to the increase in the
integrated VBF cross section from $\sigma^\mr{VBF}_\mr{NLO}=(22.5\pm
0.2)$~fb to $\sigma^\mr{VBF}_\mr{NLO+PS}=(23.9\pm 0.2)$~fb.
Similarly, the shapes of distributions related to the first tagging
jet or to the two hard leptons are found to remain stable with respect
to parton-shower effects.  This feature helps in identifying VBF
processes experimentally, as no contamination of the clean tagging-jet
signature is to be expected from parton-shower artifacts.

This statement remains true also for correlations between the leptons
and the tagging jets, such as the invariant masses of the two hardest
jets, of the two hard leptons, or angular correlations such as the
azimuthal angle separation of the tagging jets, $\Delta\phi_{j_1j_2}
=|\phi_{j_1}- \phi_{j_2}|$. The latter observable is illustrated in
Fig.~\ref{fig:phijj} (left), again at NLO and at NLO+PS level.
\begin{figure}[tp]
\begin{center}
\includegraphics[width=0.48\textwidth]{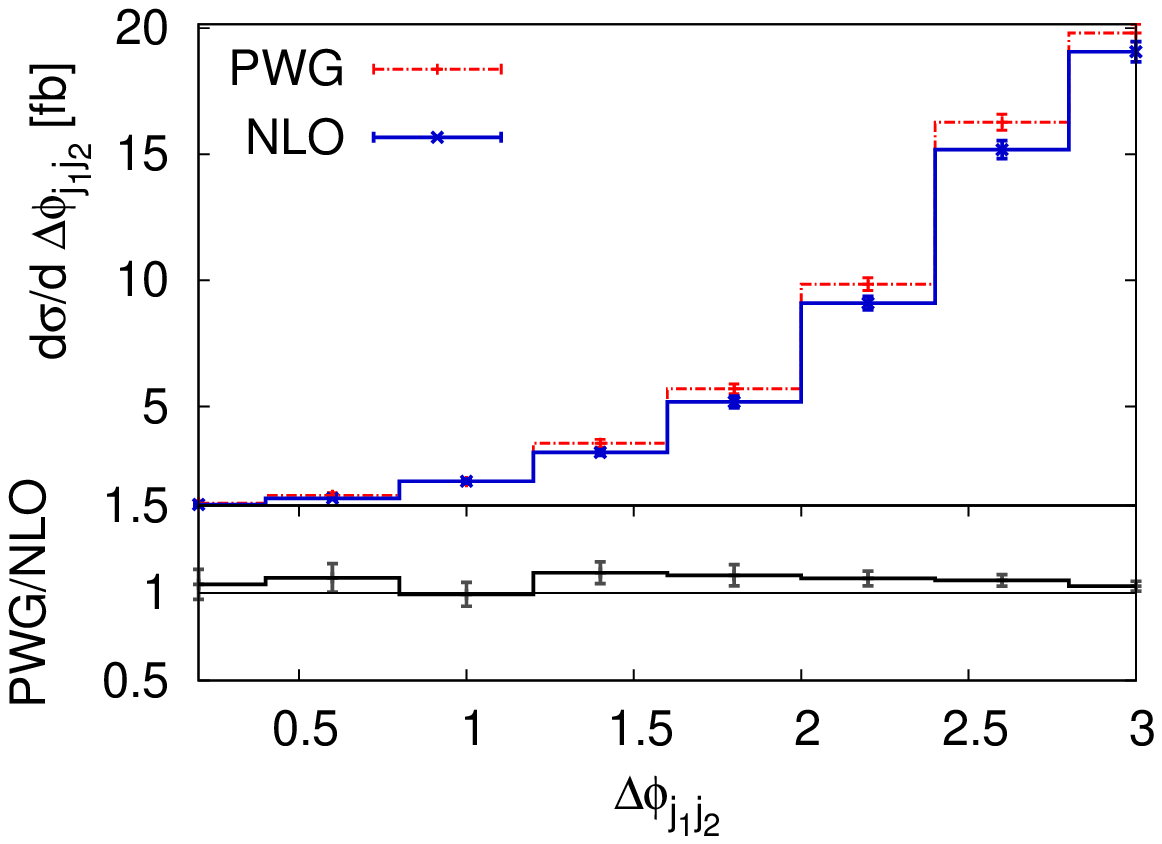}
\includegraphics[width=0.48\textwidth]{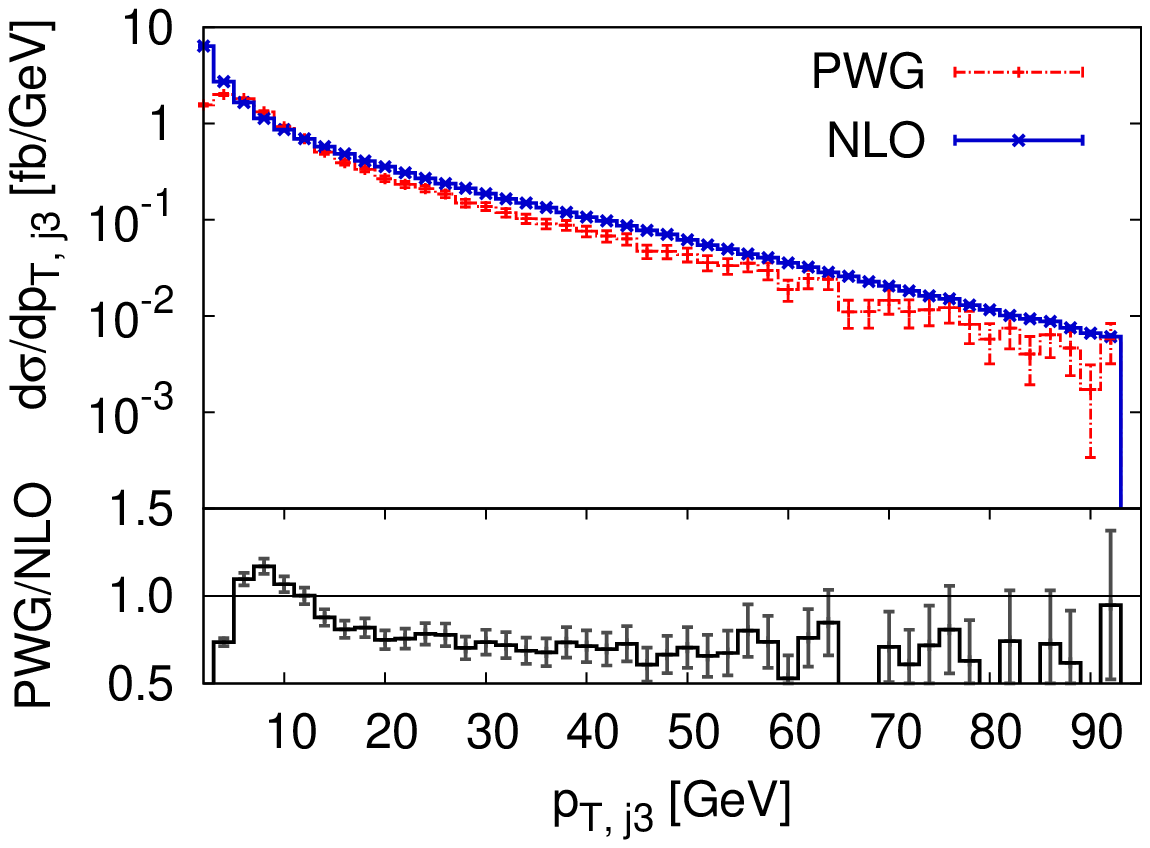}
\end{center}
\caption{
\label{fig:phijj}
As in Fig.~\ref{fig:jet2} for the azimuthal angle separation of the
two tagging jets (left panel) and for the transverse momentum distribution of the
third jet with VBF cuts (right panel). 
}
\end{figure}

More pronounced effects of the parton shower are expected for
observables that are sensitive to the emission of partons that are not
present in the LO configuration of the process under consideration. In
a fixed-order perturbative calculation for \vbfz production, at NLO
QCD a third jet can only stem from the real-emission
contributions. When the NLO calculation is merged with \PYTHIA{},
however, extra radiation can also be produced via the parton
shower. One therefore expects that the parton shower modifies
distributions of the third jet more significantly than those of the
hard tagging jets and leptons.
Figure~\ref{fig:phijj} (right) demonstrates that, indeed, the
transverse momentum distribution of the third jet (i.e., the jet of
third-highest $p_T$ that is located within the rapidity range of the
detector, $|y_j|<4.5$) changes its shape when the NLO calculation is
combined with \PYTHIA{}. While $d\sigma/dp_{T,j_3}$ in the fixed-order
NLO calculation increases steadily towards small transverse momenta,
the rise is damped by the Sudakov form factor in the {\tt
  POWHEG+PYTHIA} results.

In order to ease the identification of a jet in an experiment it is
typically required to exhibit a transverse momentum larger than
20~GeV.  Therefore, for the rapidity distribution of the third jet
shown in the left panel of Fig.~\ref{fig:y3} we considered only events
with a third jet fulfilling the requirements
\beq
\label{eq:jet3-cut}
p_{T,j_3}>20~\mr{GeV}\,,\quad
|y_{j_3}|<4.5\,.
\eeq
Apparently, the parton shower tends to fill the central-rapidity
region  slightly more than the pure NLO configuration. This effect
becomes even more pronounced, if we allow for a softer third jet,
weakening the transverse-momentum cut of Eq.~(\ref{eq:jet3-cut}) from
20 to 10~GeV, as illustrated by Fig.~\ref{fig:y3}. 
\begin{figure}[tp]
\includegraphics[width=0.5\textwidth]{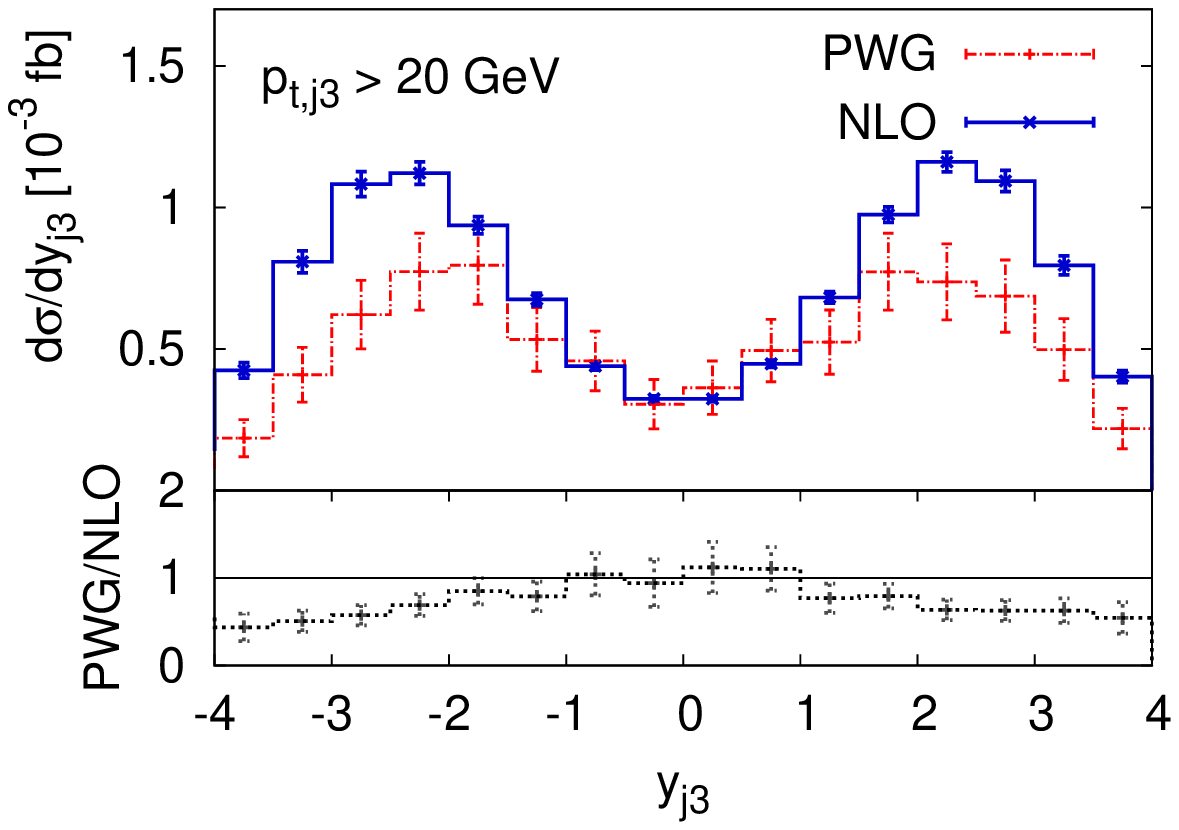}
\hfill
\includegraphics[width=0.5\textwidth]{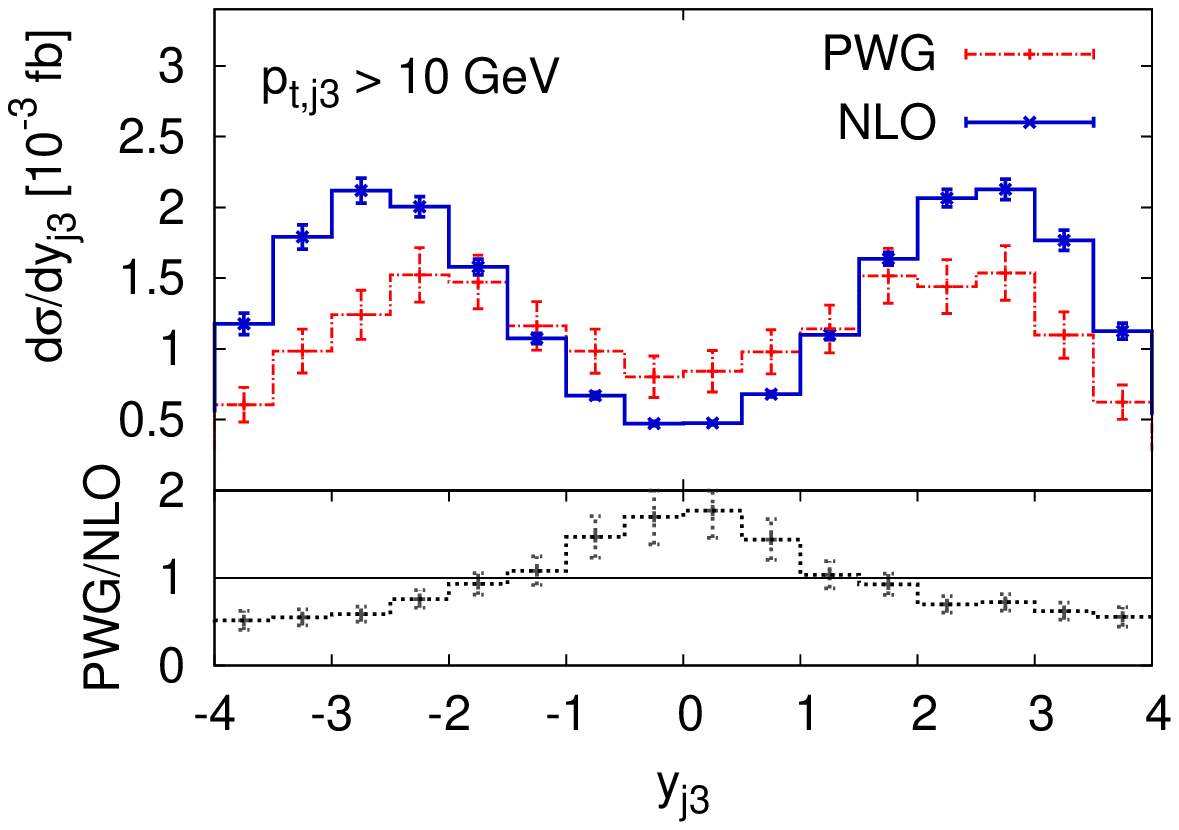}
\caption{
\label{fig:y3}
As in Fig.~\ref{fig:jet2} for the rapidity distribution of the third
jet with VBF cuts and $p_{T,j_3}>20$~GeV (left panel) and with
$p_{T,j_3}>10$~GeV (right panel).  }
\end{figure}

An observable
particularly suitable for accessing the location of the third jet in
rapidity relative to the tagging jets is the variable
\beq
y^\star = y_{j_3}-\frac{y_{j_1}+y_{j_2}}{2}\,.
\eeq
Figure~\ref{fig:ystar-vbf} 
%
%
\begin{figure}[tp]
\includegraphics[width=0.5\textwidth]{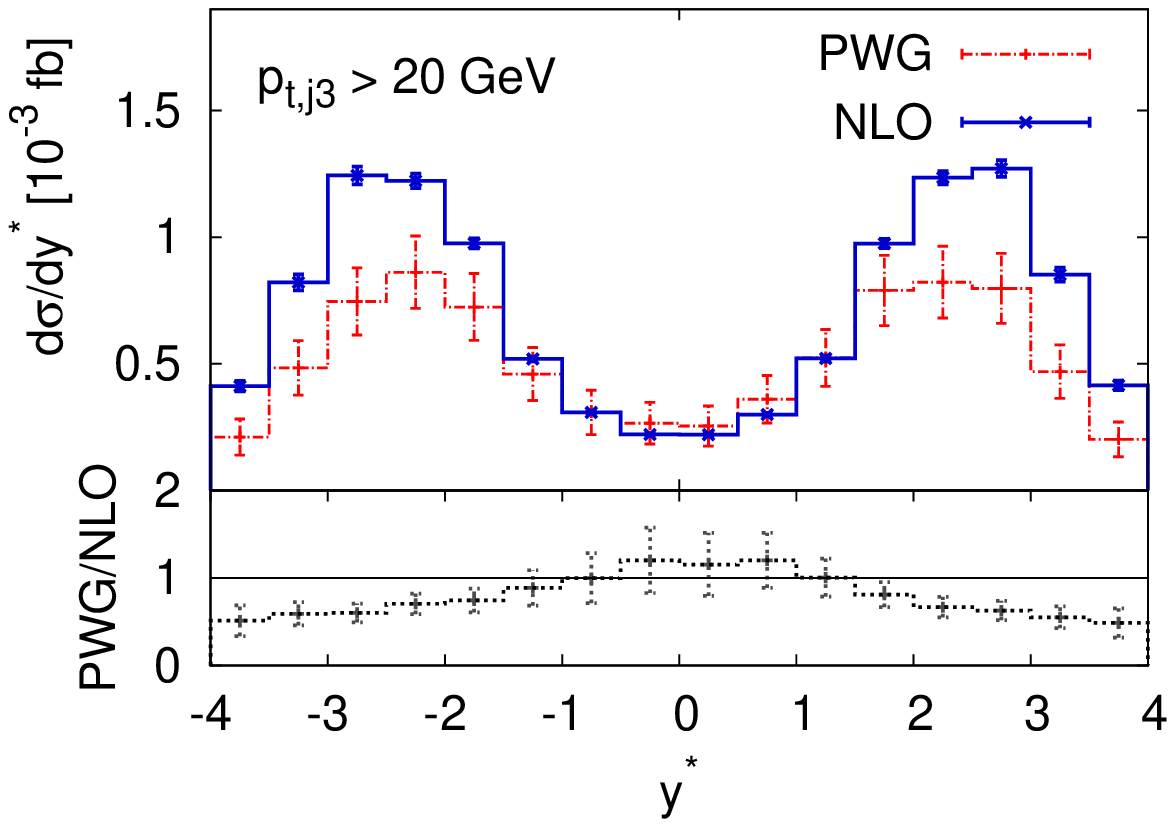}
\hfill
\includegraphics[width=0.5\textwidth]{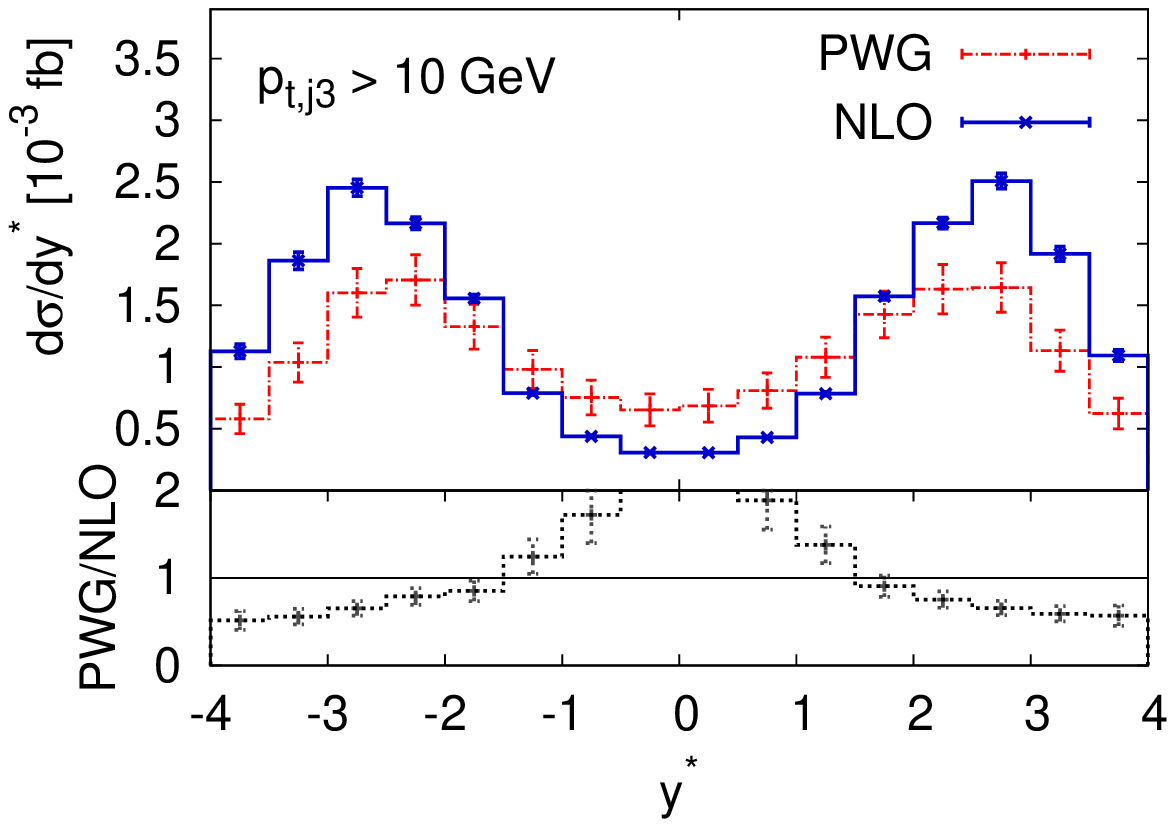}
\caption{
\label{fig:ystar-vbf}
As in Fig.~\ref{fig:jet2} for the rapidity distribution of the third
jet with respect to the average of the two tagging jets with VBF cuts
and $p_{T,j_3}>20$~GeV (left panel) and with $p_{T,j_3}>10$~GeV (right
panel).  }
\end{figure}
shows that the third jet tends to be located close to one of the
tagging jets that are peaked in the rapidity range
$|y_{j_{1,2}}|\approx 2.5\div 3$ (c.f.\ Fig.~\ref{fig:jet2}),
resulting in a maximum of the distribution slightly below
$|y^\star|\approx 3$. However, a parton that ends up close to one of
the tagging jets is likely to be recombined into this jet, giving rise
to a slight dip in $d\sigma/dy^\star$ at rapidities related to a
maximum in the distribution of a tagging jet.
Little radiation occurs in the region in the middle of the two hard
jets. If one requires the third jet to fulfill the cuts of
Eq.~(\ref{eq:jet3-cut}), the parton-shower does not change this
feature significantly. The rapidity gap is filled to some extent,
however, by softer jets, as illustrated by the plot on the
right-hand-side, where we allow for a third jet of transverse momentum
as low as 10~GeV, while all other settings remain un-altered.

A further reduction of QCD backgrounds to the VBF signal in $Zjj$
final states is expected from a central jet veto, which exploits the
unique feature of VBF reactions exhibiting two hard tagging jets that
are well separated in rapidity. In addition to the cuts of
Eqs.~(\ref{eq:pt-vbf-cut})--(\ref{eq:rapgap-cut}) we therefore explore
the impact of disregarding all events that exhibit at least one extra
jet with
\beq
\label{eq:ptveto}
p_T^\mr{veto}>20~\mr{GeV}\,,
\eeq
in the rapidity range between the two tagging jets,
\beq
\label{eq:rapveto}
\min\{y_{j_1},y_{j_2}\}<y^\mr{veto}<\max\{y_{j_1},y_{j_2}\}\,.
\eeq
In case additional jets are emitted they are expected to be located
close to the tagging jets rather than in the central-rapidity
region. While in a fixed-order parton-level calculation for \vbfz
production only one extra jet can be produced via the real-emission
contributions, in {\tt POWHEG+PYTHIA} one or more extra jets can be
produced via the parton shower. For employing CJV techniques it is
essential to understand how such parton-shower effects mitigate the
rapidity gap characteristic to VBF reactions.  We find that the
integrated cross section is reduced by roughly the same modest amount
at NLO and NLO+PS level to $\sigma^\mr{CJV}_\mr{NLO}=(19.3\pm 0.2)$~fb
and $\sigma^\mr{CJV}_\mr{NLO+PS}=(21.1\pm 0.2)$~fb. This behavior is
very different from what one expects in the case of QCD-induced $Zjj$
production, c.f.~Ref.~\cite{Re:2012zi}.

\section{Conclusion}
\label{sec:conc}
In this work we have presented an implementation of \vbfz production
in the \POWHEGBOX{}, a framework for merging NLO-QCD calculations with
parton-shower programs. We have described the technical details of our
implementation, in particular the measures taken to deal with singular
regions of the underlying Born configuration. Extensive numerical
studies have been performed to verify the independence of
phenomenological predictions on technical cuts and reweighting
factors.

We have then presented numerical results for observables that are
expected to be utilized in searches for VBF processes at the LHC. Our
analysis revealed that the parton shower may change the normalization
of cross sections by about 10~to~15~percent, but barely affect the
shapes of distributions of the two hardest jets and
leptons. Distributions related to additional jet radiation may
experience larger changes. However, the benefits of a central-jet veto
are hardly diminished by parton-shower effects.

{\bf Acknowledgments} 
We are grateful to Paolo Nason for helpful discussions and to Carlo
Oleari for a careful reading of the manuscript and valuable comments.
The work of B.~J.\ is supported in part by the Research Center {\em
  Elementary Forces and Mathematical Foundations (EMG)} of the
Johannes-Gutenberg-Universit\"at Mainz. S.~S.\ acknowledges support
from the German Research Foundation (DFG).  G.~Z.\ is supported by the
British Science and Technology Facilities Council, by the LHCPhenoNet
network under the Grant Agreement PITN-GA-2010-264564 and by the
European Research and Training Network (RTN) grant Unification in the
LHC ERA under the Agreement PITN-GA-2009-237920.

%

\end{document}